\documentclass[epj,nopacs]{svjour}
\usepackage[dvips]{graphicx} 
\usepackage{subfigure}
\newcommand\nn{\nonumber}

\newcommand{\M} {{\cal M}} 

\begin{document}

\title{Influence of Photon Inverse Emission on Forward-Backward Asymmetry in Dilepton Production at the LHC}

\author{
    V.\,A.~Zykunov\inst{1,2}
    \thanks{e-mail: zykunov@cern.ch}
}
\institute
{Joint Institute for Nuclear Research, Dubna, Russia
 \and
 Francisk Skorina Gomel State University, Belarus}
\date{Received: date / Revised version: date}

\authorrunning{V.\,A.~Zykunov}
\titlerunning{Influence of Photon Inverse Emission on Forward-Backward Asymmetry}

\abstract{
The contribution of photon inverse emission to dilepton production in hadron collisions
at the Large Hadron Collider (LHC) is calculated in detail.
Numerical analysis of inverse emission effects on cross sections
and forward-backward asymmetry is performed in a wide kinematic region 
covering the CMS experiment at the Run 3/HL-LHC regime, corresponding to 
ultra-high energies and high dilepton invariant masses.
We apply an effective technique using additive relative corrections 
to analyse the impact of radiative contributions on forward-backward asymmetry.
}

\maketitle

\section{Introduction}
\label{SectionIntroduction}

Research on lepton pair production in hadron collisions provides substantial 
opportunities to discover new physical phenomena. 
This is exemplified by the seminal discovery of the $W$- \cite{CERN-W} and $Z$-bosons \cite{CERN-Z}, 
the carriers of the weak interaction, by the UA1 and UA2 collaborations at CERN's proton collider 
during the early 1980s.
Current experiments at the LHC operate at energy scales well above the TeV threshold, 
providing a unique opportunity to search for deviations from the Standard Model (SM) 
and signatures of New Physics (NP). 
This programme encompasses diverse search channels, including: supersymmetric particles \cite{SUSY}, 
(super)string theory and $M$-theory manifestations \cite{M-theory}, 
dark matter candidates \cite{Dark-matter}, 
axions \cite{Axions}, and ultraweakly (feebly) interacting particles (FIPs) \cite{FIP}. 

The complementary nature of these searches enhances the discovery potential across 
a broad spectrum of NP scenarios.
A promising avenue involves the precise measurement of lepton pair (dilepton) production observables, 
especially in the regime of large dilepton invariant masses, 
where NP effects are expected to manifest as deviations in cross sections, 
angular distributions, and forward-back\-ward asymmetry. 
Careful examination of these observables at the LHC thus offers a sensitive probe 
for physics beyond the Standard Model.

The Drell--Yan process \cite{DY2,DY2-2}, defined as the production of a dilepton in hadron collisions 
via the annihilation of a quark-antiquark pair through a virtual photon or $Z$-boson, 
is conventionally distinguished from other dilepton production processes 
that also involve hadron collisions but proceed via different mechanisms.
Beyond the Drell--Yan mechanism, several alternative dilepton production channels 
must be considered in hadron collisions. 
One notable example is photon fusion, in which quasi-real photons emitted 
from the incoming hadrons interact to produce a dilepton pair. 
Another important class of processes involves boson inverse emission -- 
specifically, gluon or photon radiation from an initial-state parton -- 
which can also lead to dilepton final states. 
These mechanisms often contribute as higher-order corrections or background processes 
in precision measurements at the LHC.

Most probably, NP will manifest itself in LHC experiments not via direct detection of new particles
and/or new phenomena but in the form of minor deviations from SM predictions. 
The discovery of new physical phenomena in such a scenario is only possible
by an extremely precise comparison of the experimental data with theoretical predictions. 
This requires the corresponding increasing of the accuracy of the theoretical description 
of the microworld processes under study and provision of experiments 
to be conducted at Run3/HL LHC with reliable and precision programs
for calculating effects related to radiative corrections (RCs).

Currently, a large number of various and mutually complementing programs and computer codes are 
available that pertain to this problem; they are reviewed, for example, in \cite{50-Au}. 
This paper also presents the physical content of one of these programs, 
READY (Radiative corrEctions to lArge invariant mass Drell--Yan process), which was developed by the 
author for estimating electroweak and QCD corrections to the Drell--Yan process.
The contribution of the photon inverse emission mechanism is of the same order in perturbation theory as 
the electroweak RCs to the Drell--Yan process or the two-photon fusion mechanism. 
Therefore, the goal of the present paper is the detailed calculation of the photon inverse emission contribution 
to the dilepton production process in hadron collisions, 
aimed at the Large Hadron Collider experimental program.

\section{Mechanisms of Dilepton Production in Hadron Collisions}

\begin{figure}
    \centering
    \mbox{
        \subfigure[]{\includegraphics[width=0.18\textwidth]{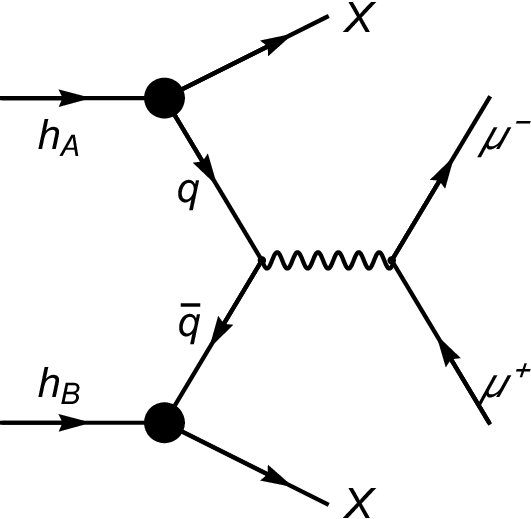}\label{F1}}
        \quad
        \subfigure[]{\includegraphics[width=0.18\textwidth]{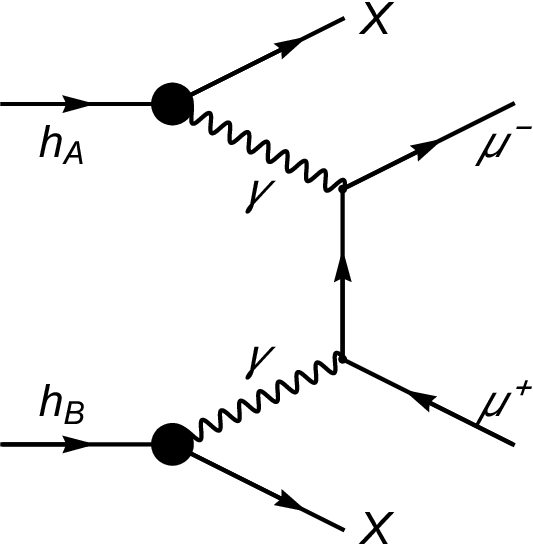}\label{F2}}
    }
    \\
    \mbox{
        \subfigure[]{\includegraphics[width=0.18\textwidth]{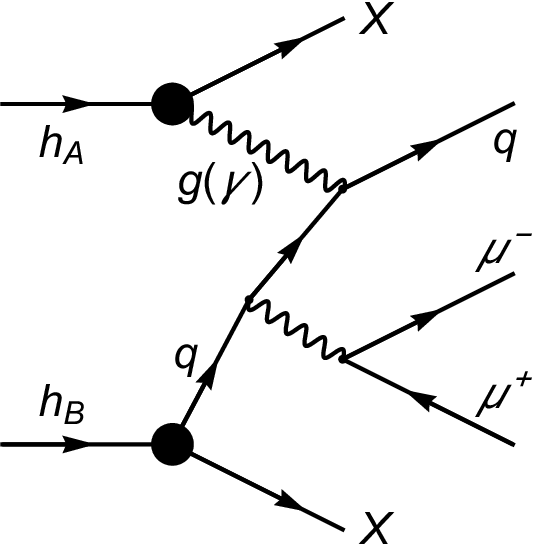}\label{F3}}
        \quad
        \subfigure[]{\includegraphics[width=0.18\textwidth]{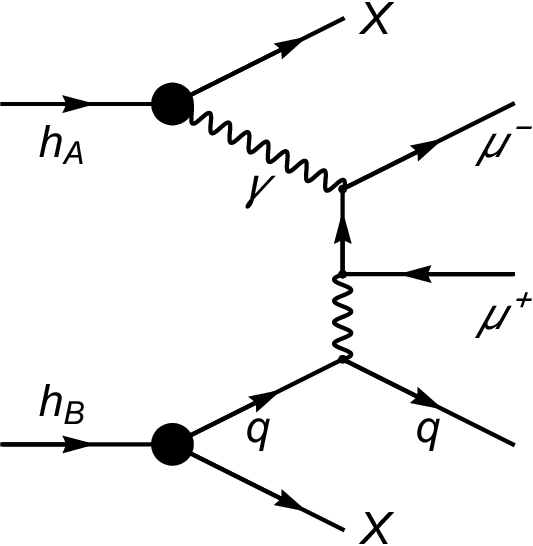}\label{F4}}
    }
    \caption{
Dimuon production mechanisms in hadron collisions:
{a} -- Drell--Yan, 
{b} -- $\gamma\gamma$ fusion,
{c} -- quark-initiated photon/gluon inverse emission,
{d} -- lepton-initiated photon inverse emission.
Unsigned lines: $\gamma$ or $Z$-boson.
}
    \label{fig.hh}
\end{figure}

There are different mechanisms of dilepton ($\ell^-\ell^+,\ \ell = e,\mu,\tau$) production in hadron collisions:
\begin{equation}
h_A(p_A) + h_B(p_B) \rightarrow \ell^-(p_3) + \ell^+(p_4) + X,
\label{hh-mumu}
\end{equation}
where
$p_A$ and $p_B$ are the 4-momenta of the initial hadrons;
$p_3$ $(p_4)$ is the 4-momentum of the final lepton $\ell^-$ (antilepton $\ell^+$) with mass $m$.
Three possibilities are illustrated in Fig.~\ref{fig.hh}, with the discussion primarily 
focused on the dimuon case $\ell=\mu$.

First, we consider the $s$-channel mechanism of quark-antiquark annihilation into a dimuon pair, 
as illustrated in Fig.~\ref{fig.hh}(a).
The Drell--Yan process, which is realized via this mechanism, 
is currently well studied, including effects at the level of one-loop electroweak radiative corrections 
(RCs) and two-loop quantum chromodynamics (QCD) RCs (see, for example, \cite{50-Au} and the references therein).
The second possibility is $\gamma\gamma$ fusion [see Fig.~\ref{fig.hh}(b)]. 
This mechanism has been investigated up to the level of one-loop electroweak RCs, 
as reported in \cite{55-Au,56-Au}.
To achieve the accuracy of order ${\cal O} (\alpha^3)$, we must take into account 
both gluon inverse emission (see \cite{22-Au} and the recent work \cite{60-Au}) 
and photon inverse emission, as illustrated in Fig.~\ref{fig.hh}(c) and Fig. \ref{fig.hh}(d).
Below, we will denote these mechanisms and their parton-level configurations as follows:
\begin{enumerate}
\item
 DY = $\{ q \bar q, \bar q q\}$ -- the Drell--Yan process;
\item
$\gamma\gamma$ -- two-photon fusion;
\item
$g\rm IE$ = $\{ gq, qg, g\bar q, \bar q g\}$ -- gluon inverse emission;
\item
$\gamma\rm IE$ = $\{ \gamma q, q \gamma, \gamma \bar q, \bar q \gamma\}$ -- photon inverse emission.
\end{enumerate}
The first symbol within the curly braces denotes the parton originating from the first hadron, $h_A$, 
while the second symbol corresponds to the parton from the second hadron, $h_B$.
The parton mechanisms  $q\bar q, gq, g\bar q, \gamma q, \gamma \bar q$
will be referred to as {\it direct configurations}, while the mechanisms
$\bar q q, q g, \bar q g, q \gamma,\bar q \gamma$
will be termed {\it reverse configurations}.
For the observables considered in this paper, 
both configurations yield identical contributions. 
Additionally, we introduce a common index $c$ to denote the type of mechanism: 
\begin{equation}
c = {\rm DY}, \gamma\gamma, g {\rm IE}, \gamma {\rm IE}.
\label{c}
\end{equation}

\section{Parton-Level Amplitudes}

\begin{figure}
    \centering
    \mbox{
        \subfigure[]{\includegraphics[width=0.18\textwidth]{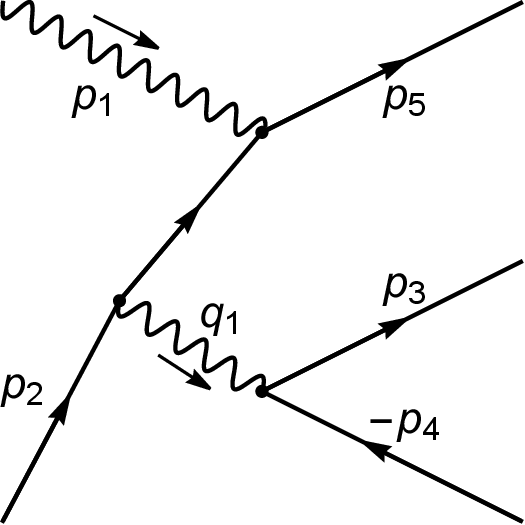}\label{F5}}
        \quad
        \subfigure[]{\includegraphics[width=0.18\textwidth]{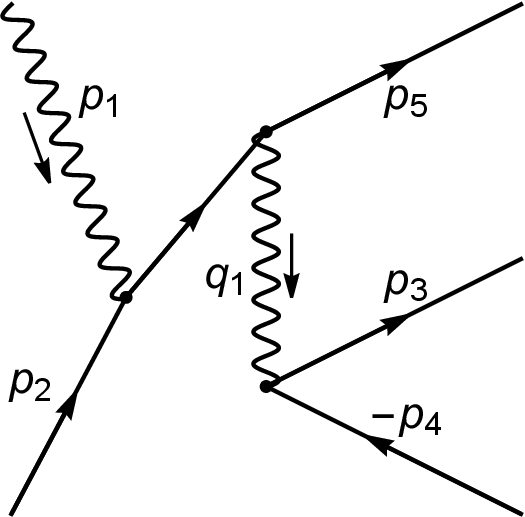}\label{F6}}
    }
    \\
    \mbox{
        \subfigure[]{\includegraphics[width=0.18\textwidth]{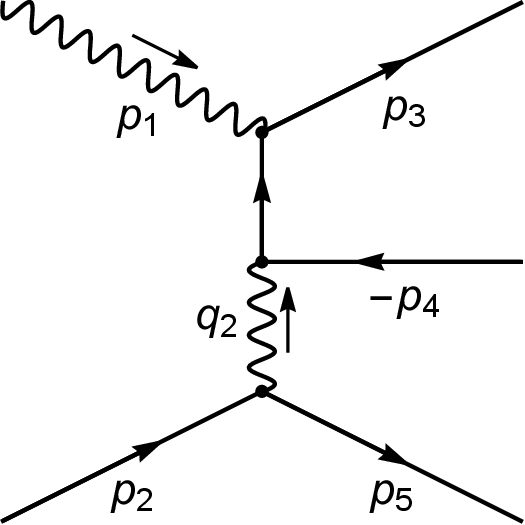}\label{F7}}
        \quad
        \subfigure[]{\includegraphics[width=0.18\textwidth]{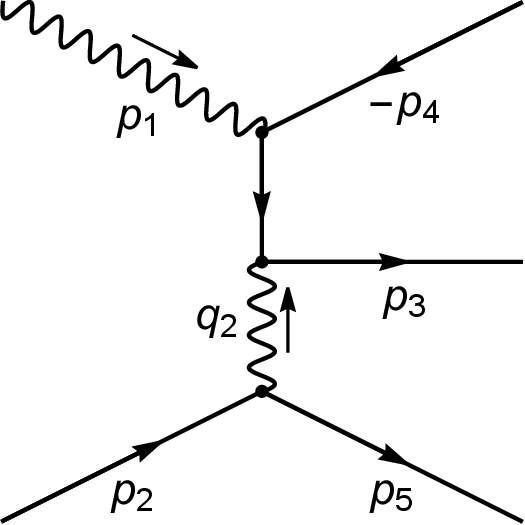}\label{F8}}
    }
\caption{ 
Feynman diagrams for photon inverse emission: 
upper diagrams show photon-quark interactions, which can easily be converted into gluon-quark diagrams.
}
\label{fig.pq} 
\end{figure}

At the parton level in the direct configuration, the dilepton production process 
with photon inverse emission takes the following form:
\begin{equation}
\gamma(p_1) + q(p_2) \rightarrow \ell^-(p_3) + \ell^+(p_4) + q(p_5).
\label{ige}
\end{equation}
The Feynman diagrams corresponding to process (\ref{ige}) are presented in Fig.~\ref{fig.pq}.

The notations used in Fig.~\ref{fig.pq} are defined as follows:
$p_1$ denotes the 4-momentum of the initial photon;
$p_2\ (p_5)$ denotes the 4-momentum of the initial (final) quark;
$q_1=p_3+p_4$ or $q_2=p_2-p_5$ represent the 4-momenta of the intermediate boson (photon or $Z$-boson) 
of type $a$ and $b$, respectively, where $a,b=\{\gamma,Z\}$;
$p_3\ (p_4)$ denotes the 4-momentum of the lepton (antilepton).
Let us introduce the six partonic Lorentz invariants, which are defined as follows:
\begin{eqnarray}
&& s = (p_1+p_2)^2,\ \ t = (p_1-p_3)^2,\ \ u = (p_2-p_3)^2,
\label{part-inv}
\\
&& s_1 = (p_3+p_4)^2,\ \ t_1 = (p_2-p_4)^2,\ \ u_1 = (p_1-p_4)^2.
\nn
\end{eqnarray}
Additionally, we will employ the following Lorentz invariants:
\begin{equation}
 w_1=2p_1p_5,\ \ w_2=2p_2p_5,\ \ w_3=2p_3p_5,\ \ w_4=2p_4p_5.
\label{rad-invariants-qq}
\end{equation}

Let us briefly present the Feynman rules employed in our calculations, 
which are based on the framework described in \cite{BSH86}:
\begin{enumerate}
\item
An incoming fermion with 4-momentum $p$ is represented by 
the bispinor $u(p)$, while an outgoing fermion is described by the bispinor $\bar u(p)$.
\item
An incoming boson with 4-momentum $p$ is represented by the polarization vector $e_\rho(p)$.
\item
The boson propagator is expressed as
$- i g^{\alpha\beta} D_{a}(q)$,
where $q$ is the 4-momentum transferred through the propagator, and
\begin{equation}
D_{a}(q)=(q^2-m_a^2+im_a\Gamma_a)^{-1}.
\nn
\end{equation}
Photon mass is zero, 
mass of $Z$-boson denoted by $m_Z$, and its fixed decay width is $\Gamma_Z$.
\item
The fermionic propagator for a particle with 4-momen\-tum $p$ is given by the expression:
\begin{eqnarray}
i S(p) = i \frac{\hat p + m_f}{p^2-m_f^2},\ \ \ \hat p \equiv \gamma^\mu p_\mu.
\nn
\end{eqnarray}
\item
The vertex describing the interaction between a fermion $f$ and a gauge boson $a$ 
is represented by the expression
\begin{equation}
 i e \gamma_\mu \Gamma^a_f,\ \ \mbox{where}\ \   
\Gamma^a_f  = v^a_f - a^a_f \gamma_5,
\nn
\end{equation} 
and vector and axial-vector coupling constants have the following form:
\begin{equation}
\!\!\!\!\!
v_f^{\gamma}=-Q_f,\ \ a_f^{\gamma}=0,\ \
v_f^Z=\frac{I_f^3-2Q_fs_{\rm W}^2}{2s_{\rm W}c_{\rm W}},\ \ 
a_f^Z=\frac{I_f^3}{2s_{\rm W}c_{\rm W}}.
\nn
\end{equation}

\end{enumerate}

We used the following SM parameters:
$Q_f$ is the electric charge of the $f$-particle in units of the proton charge ($e$);
the third components of weak isospin are:
\begin{equation}
I_{\nu}^3=+\frac{1}{2},\ \
I_\ell^3 =-\frac{1}{2},\ \
I_u^3    =+\frac{1}{2},\ \
I_d^3    =-\frac{1}{2},
\nn
\end{equation}
and the sine $s_{\rm W}$ and cosine $c_{\rm W}$ of the weak mixing angle are expressed 
in terms of the masses of the $Z$- and $W$-bosons according to the SM relations:
\begin{equation}
c_{\rm W}=\frac{m_W}{m_{Z}},\ \ \
s_{\rm W}=\sqrt{1-c_{\rm W}^2}.
\nn
\end{equation}

Using the aforementioned Feynman rules, we can write the amplitude 
of the $\gamma q$-process with quark interaction and an intermediate $a$-boson, 
as represented by the diagrams in Fig.~\ref{fig.pq}(a,b):
\begin{eqnarray}
\M_{q}^a &=& ie^3Q_q D_a(q_1) e^\mu(p_1)
 \cdot   \bar u(p_5) \bigl[ \gamma_\mu S(p_5-p_1) \gamma_\alpha \Gamma_q^a +
\nn \\
&+& \gamma_\alpha \Gamma_q^a S(p_1+p_2) \gamma_\mu \bigr] u(p_2) 
 \cdot \bar u(p_3) \gamma^\alpha \Gamma_\ell^a u(-p_4) .
\label{Agq}
\end{eqnarray}
The Dirac equation and the commutation rules for the $\gamma$-matrices 
enable the simplification of the amplitude (\ref{Agq}):
\begin{eqnarray}
\M_{q}^a &=& ie^3Q_q D_a(q_1) e^\mu(p_1) \times
\nn \\
 &\times & \bar u(p_5) G^{q}_{\mu \alpha} \Gamma_q^a u(p_2) 
 \cdot \bar u(p_3) \gamma^\alpha \Gamma_\ell^a u(-p_4),
\label{Agq2}
\end{eqnarray}
where 
\begin{equation}
G^{q}_{\mu \alpha} =
   \frac{2 {p_5}_\mu - \gamma_\mu \hat p_1}{-2p_1p_5} \gamma_\alpha
+  \gamma_\alpha \frac{\hat p_1 \gamma_\mu + 2 {p_2}_\mu }{2p_1p_2}.
\nonumber
\end{equation}
Similarly, the amplitude of the $\gamma q$-process with lepton interaction, 
as depicted in the diagrams of Fig.~\ref{fig.pq},(c,d), takes the following form:
\begin{eqnarray}
\M_{\ell}^a &=& ie^3 Q_\ell D_a(q_2) e^\mu(p_1) \times
\nn \\
 &\times & \bar u(p_3) G^{\ell}_{\mu \alpha} \Gamma_\ell^a u(-p_4) 
 \cdot \bar u(p_5) \gamma^\alpha \Gamma_q^a u(p_2),
\label{Agl2}
\end{eqnarray}
where 
\begin{equation}
G^{\ell}_{\mu \alpha} =
   \frac{2 {p_3}_\mu - \gamma_\mu \hat p_1}{-2p_1p_3} \gamma_\alpha
+  \gamma_\alpha \frac{\hat p_1 \gamma_\mu - 2 {p_4}_\mu }{-2p_1p_4}.
\nn
\end{equation}

\section{Squaring at the Parton Level}

Applying Casimir's trick and averaging over the polarizations of the initial particles, 
as well as summing over the polarizations of the final particles, yields the squared amplitude (\ref{Agq2}):
\begin{equation}
\overline{\sum\limits_{\rm pol}} \M_{q}^a (\M_{q}^b)^+ 
=
e^6 Q_q^2 D_a(q_1) D_b^*(q_1)\, S^{ab}_{q}, 
\label{MM}
\end{equation}
where
\begin{equation}
S^{ab}_{q} =
- \frac{1}{4} 
{\rm Sp} \bigl[ G^{q}_{\mu \alpha} G_q^{ab} \hat p_2 {G^{q}_{\mu \beta}}^+ \hat p_5 \bigr]
{\rm Sp} \bigl[ \gamma^\alpha G_l^{ab} \hat p_4 \gamma^\beta \hat p_3 \bigr].
\label{Spur-part-q}
\end{equation}
Where applicable, the ultrarelativistic approximation 
\begin{equation}
m_f \ll \sqrt{s}
\label{ua}
\end{equation}
has been employed.
Additionally, we have utilized the formal property for the polarization vectors of photons:
$${\sum_{\rm pol}}  e_\rho(p)  e_{\rho'}(p)  = - g_{\rho \rho'}.$$
The coupling constants in (\ref{Spur-part-q}) are factorized as follows:
\begin{equation}
G_f^{ab} = {\Gamma_f^a} {\Gamma_f^b} = \lambda^{ab}_{fV} - \lambda^{ab}_{fA}\gamma_5,\ \ \
f=q,\ell,
\nn
\end{equation}
where the expressions for $\lambda$ assume rather simple forms:
\begin{equation}
\lambda^{ab}_{fV} = v_f^a v_f^b + a_f^a a_f^b,\ \ \
\lambda^{ab}_{fA} = v_f^a a_f^b + a_f^a v_f^b.
\nn
\end{equation}

Squaring (\ref{Agl2}) by applying the same techniques, we obtain:
\begin{equation}
\overline{\sum\limits_{\rm pol}} \M_{\ell}^a (\M_{\ell}^b)^+ =
e^6 Q_\ell^2 D_a(q_2) D_b^*(q_2)\, S^{ab}_{\ell}, 
\label{MM-l}
\end{equation}
where
\begin{equation}
S^{ab}_{\ell} =
- \frac{1}{4} 
{\rm Sp} \bigl[ G^{\ell}_{\mu \alpha} G_\ell^{ab} \hat p_4 {G^{\ell}_{\mu \beta}}^+ \hat p_3 \bigr]
{\rm Sp} \bigl[ \gamma^\alpha G_q^{ab} \hat p_2 \gamma^\beta \hat p_5 \bigr].
\label{Spur-part-l}
\end{equation}
Finally, the interference between the $q$- and $\ell$-cases takes the following form:
\begin{eqnarray}
\overline{\sum\limits_{\rm pol}} \M_{q}^a (\M_{\ell}^b)^+ 
&=& e^6 Q_qQ_\ell D_a(q_1) D_b^*(q_2)\, S^{ab}_{q\ell}, 
\nn\\
\overline{\sum\limits_{\rm pol}} \M_{\ell}^a (\M_{q}^b)^+ 
&=& e^6 Q_qQ_\ell D_a(q_2) D_b^*(q_1)\, S^{ab}_{\ell q}, 
\label{MM-ql}
\end{eqnarray}
where
\begin{eqnarray}
\!\!\!\! S^{ab}_{q\ell} &=&
- \frac{1}{4} 
{\rm Sp} \bigl[ G^{q}_{\mu \alpha} G_q^{ab} \hat p_2 \gamma^\beta  \hat p_5 \bigr]
{\rm Sp} \bigl[ \gamma^\alpha G_\ell^{ab} \hat p_4 {G^{\ell}_{\mu \beta}}^+ \hat p_3 \bigr],
\label{Spur-part-ql}
\\
\!\!\!\! S^{ab}_{\ell q} &=&
- \frac{1}{4} 
{\rm Sp} \bigl[ G^{\ell}_{\mu \alpha} G_\ell^{ab} \hat p_4 \gamma^\beta  \hat p_3 \bigr]
{\rm Sp} \bigl[ \gamma^\alpha G_q^{ab} \hat p_2 {G^{q}_{\mu \beta}}^+ \hat p_5 \bigr].
\label{Spur-part-lq}
\end{eqnarray}

Upon evaluating the expressions (\ref{Spur-part-q}), (\ref{Spur-part-l}), 
(\ref{Spur-part-ql}), and (\ref{Spur-part-lq}) -- using computer algebra methods 
(e.g., FORM \cite{Kuipers:2012rf} or FeynCalc \cite{FeynCalc}) -- 
we obtain, after a series of simplifications, only two combinations of coupling constants:
\begin{equation}
S^{ab}_{j} =
  \lambda_{qV}^{ab}\lambda_{\ell V}^{ab} \, S^V_{j} 
+ \lambda_{qA}^{ab}\lambda_{\ell A}^{ab} \, S^A_{j},\ \ \
j=(q,\ell,q\ell,\ell q).
\label{Spur-part_2}
\end{equation}
Here, the factors $S^{V,A}_j$
are expressed solely in terms of scalar products of 4-momenta and particle masses.

\section{Partonic and Hadronic Cross Sections}
\label{2-3}

\begin{figure}
    \centering
\scalebox{0.7}{\includegraphics{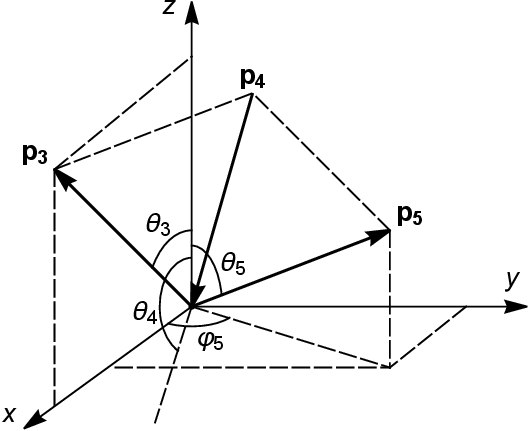}} 
\caption{ 
The configuration of the three-momenta of the final particles in the partonic centre-of-mass system.
}
\label{fig.vecs} 
\end{figure}

The process $2 \rightarrow 3$ is characterised by the following differential (partonic-level) cross section:
\begin{equation}
d\sigma_{2 \rightarrow 3} =\frac{1}{2^6 \pi^5 s} \overline{\sum_{\rm pol}}  |\M|^2 d^5\Phi_3,
\label{cs23}
\end{equation}
where the three-particle phase space $d\Phi_3$ is defined as
\begin{equation}
d^5\Phi_3 = \delta (p_1+p_2-p_3-p_4-p_5) 
\frac{d^3 {\bf p_3}}{2 E_3 } \frac{d^3 {\bf p_4}}{2 E_4} \frac{d^3 {\bf p_5}}{2 E_5}.
\label{Phi_3}
\end{equation}
The configuration of the three-momenta of the final particles 
in the partonic centre-of-mass system (c.m.s.) is illustrated in Fig.~\ref{fig.vecs}.

In the partonic centre-of-mass system, following the integrations 
facilitated by the $\delta$-function over ${\bf p_4}$ and the trivial integration over $\varphi_3$ 
(which is rendered straightforward by the rotational symmetry of the system about this axis), 
the phase space takes the following form:
\begin{equation}
d^4\Phi_3 = \frac{\pi |{\bf p_3}| |{\bf p_5}|}{4 E_4 {\cal F}} d(\cos\theta_3) dE_5 d(\cos\theta_5) d\varphi_5,
\label{Phi_3-3}
\end{equation}
where
\begin{equation}
{\cal F} = 1 + E_3 \bigl( 1+{\cal B}/|{\bf p_3}| \bigr) 
                   \bigl({\cal A}+E_3^2+2{\cal B}|{\bf p_3}| \bigr)^{-\frac{1}{2}}.
\label{F}
\end{equation}
The energy $E_3$ can be calculated as follows:
\begin{equation}
E_3 = \frac{{\cal C}({\cal A}-{\cal C}^2) + {\cal B}\sqrt{({\cal A}-{\cal C}^2)^2
            +4m_3^2 ({\cal B}^2-{\cal C}^2)}}{2({\cal B}^2-{\cal C}^2)},
\label{E3}
\end{equation}
where 
\begin{equation}
{\cal A} = m_4^2 - m_3^2 + |{\bf p_5}|^2,\ \ \
{\cal B} = |{\bf p_5}|\cos\theta_{35},\ \ \
{\cal C} = E_1+E_2-E_5.
\label{ABCDEF}
\end{equation}
Here, $\theta_{35}$ denotes the angle between $\bf p_3$ and $\bf p_5$, which is defined by the relation
$$\cos\theta_{35} = \cos\theta_3 \cos\theta_5 + \sin\theta_3 \sin\theta_5 \cos\varphi_5.$$
Using equations (\ref{cs23}) and (\ref{Phi_3-3}), we express the $j$-contribu\-ti\-on to 
the differential cross section [where $j$ is defined in equation (\ref{Spur-part_2})] as follows:
\begin{equation}
d^4\sigma_{j} =
\frac{1}{2^6 \pi^5 s}
\sum_{a,b} \overline{\sum\limits_{\rm pol}} \M_{q,\ell}^a (\M_{q,\ell}^b)^+
d^4\Phi_3.
\label{cs_ige}
\end{equation}

Based on the expression (\ref{cs_ige}), we construct the $j$-con\-tri\-bu\-tion 
to the differential hadronic cross section by applying the convolution procedure:
\begin{equation}
d\sigma_{j}^{\rm ex} =
\sum_{q}
\sum_{r_1,r_2}
f_\gamma^{r_1,A}(x_1,Q^2) dx_1 \, f_q^{r_2,B}(x_2,Q^2) dx_2\,
d^4\sigma_{j}.
\nonumber
\end{equation}
The convolution incorporates the summation over all possible dilepton production configurations, 
explicitly excluding ones that violate conservation laws.
The summation over $q$ encompasses all six quark flavours ($q=u,d,s,c,b,t$); 
however, it is customary to exclude the top quark ($t$) from the calculation due to its significantly large mass.

The symbol $f_q^{r,h}(x,Q^2)$ denotes {\it the partonic distribution function}, 
such that $f_q^{r,h}(x,Q^2) dx$ represents the probability of finding a quark of flavour $q$ 
and helicity $r$ within hadron $h$, carrying a momentum fraction in the interval from $x$ to $x+dx$, 
at the resolution scale of the reaction $Q^2$.

For brevity, we henceforth omit the dependence on $Q^2$ in the notation of $f_q$.
For process (\ref{hh-mumu}), the natural choice for the scale $Q^2$ is given by the relation:
$Q^2 = M^2$,
where $M$ denotes the invariant mass of the produced muon dilepton:
\begin{equation}
M = \sqrt{(p_3+p_4)^2}.
\label{M}
\end{equation}

We should also multiply the integrand by a factor $\Theta$, 
where $\Theta$ is the Heaviside step function (often denoted as $\theta$-function).
This function acts as a mathematical switch:
It equals 1 at integration points that correspond to physically allowed configurations -- 
i.e., those satisfying all relevant conservation laws, kinematic constraints, 
and experimental acceptance criteria.
It equals 0 at points violating any of these conditions, 
effectively excluding unphysical or inaccessible regions from the integration domain.
The inclusion of $\Theta$ in the integrand implements what is commonly 
referred to as  the {\it fiducial cuts} procedure.

Supposing the initial hadrons are unpolarized, we sum over helicities and introduce 
unpolarized parton distribution functions:
\begin{equation}
f_q^{A}(x) =
\sum_{r} f_q^{r,h}(x) = 
f_q^{+,h}(x) + f_q^{-,h}(x).
\label{unp_pdf}
\end{equation}
With this definition, the hadronic differential cross section 
for process (\ref{hh-mumu}) takes the following form:
\begin{equation}
d\sigma_{j}^{\rm ex} =
\sum_{q}
f_\gamma^{A}(x_1) f_q^{B}(x_2) \, dx_1 dx_2 \,
d\sigma_{j}.
\label{cs_hh-mumu-2}
        \end{equation}

For hadronic Lorentz invariants, we choose expressions analogous to (\ref{part-inv}), 
but with ``hats'':
%
\begin{eqnarray}
&& \hat s = (p_A+p_B)^2,\ \ \hat s_1 = (p_3+p_4)^2,
\nn\\
&& \hat t = (p_A-p_3)^2,\ \ \hat t_1 = (p_B-p_4)^2,
\label{hadr-inv} \\
&& \hat u = (p_B-p_3)^2,\ \ \hat u_1 = (p_A-p_4)^2.
\nn
\end{eqnarray}
It can be observed that $ \hat s_1 \equiv s_1 \equiv M^2$, and $ \hat s \equiv S$.

According to the quark-parton model, the 4-momenta of the hadron and parton 
are proportional, as expressed by the following relations:
\begin{equation}
p_1 = x_1 p_A,\ \ \
p_2 = x_2 p_B,
\label{QPM}
\end{equation}
where the coefficients $x_1$ and $x_2$ represent the momentum fractions 
carried by the parton (photon or quark) with respect to its parent hadron.
Then, in the ultrarelativistic approximation, the relations between the parton 
and hadron invariants take the following form:
\begin{equation}
\hat s = \frac{s}{x_1x_2},\ \ \hat t  = \frac{t}{x_1},\ \ \hat u = \frac{u}{x_2},\ \
\hat t_1 = \frac{t_1}{x_2},\ \ \hat u_1 = \frac{u_1}{x_1}.
\label{hat-s1t1u1-dy}
\end{equation}

We are now prepared to proceed with the analysis using the experimentally 
accessible variables: the invariant dilepton mass $M$ [see equation~(\ref{M})]
and the dilepton rapidity $y$:
\begin{equation}
y = \frac{1}{2} \log\frac{E+p_z}{E-p_z}
= \frac{1}{2} \log\frac{\hat t_1 + \hat u}{\hat t + \hat u_1}.
\label{rapidity}
\end{equation}
By incorporating the two particularly useful relations:
$$t_1 + u + s_1 + w_1 = 0, \ \ \ t + u_1 + s_1 + w_2 = 0,$$ 
which can be derived from the 4-momentum conservation law for a $2 \rightarrow 3$ reaction, we obtain:
\begin{equation}
y =  \frac{1}{2} \log\frac{x_1 (w_1+s_1)}{x_2 (w_2+s_1)}.
\label{rapidity1}
\end{equation}
From the equality $s-s_1-w_1-w_2=0$ in the partonic centre-of-mass system, it follows that:
\begin{equation}
\sqrt{s} =  E_5 + \sqrt{E_5^2+s_1}.
\label{sqs}
\end{equation}

Upon solving the system of equations (\ref{hat-s1t1u1-dy}), (\ref{rapidity1}), and (\ref{sqs}), we obtain:
\begin{eqnarray}
x_1 &=& \frac{E_5 + \sqrt{E_5^2+M^2}}{\sqrt{S}} \sqrt{\frac{w_2+M^2}{w_1+M^2}} \, e^{y},
\label{x1-x2-R}
\\ 
x_2 &=& \frac{E_5 + \sqrt{E_5^2+M^2}}{\sqrt{S}} \sqrt{\frac{w_1+M^2}{w_2+M^2}} \, e^{-y}.
\nn
\end{eqnarray}
Taking into account the aforementioned expressions, we can proceed to the variables $M$ and $y$ as follows:
\begin{equation}
dx_1 dx_2 
= J_x \, dM dy,\ \ \
J_x = \frac{ 2M (E_5 + \sqrt{E_5^2+M^2}) }{ S\sqrt{E_5^2+M^2} }.
\label{Jx}
\end{equation}
Finally, substituting the expressions (\ref{MM}), (\ref{cs_ige}), and (\ref{Jx}) 
into equation~(\ref{cs_hh-mumu-2}) yields the $j$-component of the differential 
cross section for the process (\ref{hh-mumu}):
\begin{eqnarray}
d\sigma_{j}^{\rm ex} &=&
\frac{\alpha^3 J_x}{\pi^2 s}
\sum_{q}
f_\gamma^{A}(x_1) f_q^{B}(x_2) 
\times 
\label{cs_hh-mumu-3}
\\
&\times &
\bigl[ V_{j}\, S^V_j + A_{j}\, S^A_j \bigr] \, d^4\Phi_3\, dM dy,
\nn
\end{eqnarray}
where the vector combinations are defined as follows:
\begin{eqnarray} 
V_{q} &=& Q_q^2 \sum\limits_{a,b=\gamma, Z} \lambda_{qV}^{ab}\lambda_{\ell V}^{ab} D_a(q_1)D_b^*(q_1),
\nn\\
V_{\ell} &=& Q_\ell^2 \sum\limits_{a,b=\gamma, Z} \lambda_{qV}^{ab}\lambda_{\ell V}^{ab} D_a(q_2)D_b^*(q_2),
\nn\\
V_{q\ell} &=& Q_qQ_\ell \sum\limits_{a,b=\gamma, Z} \lambda_{qV}^{ab}\lambda_{\ell V}^{ab} D_a(q_1)D_b^*(q_2),
\nn\\
V_{\ell q} &=& Q_qQ_\ell \sum\limits_{a,b=\gamma, Z} \lambda_{qV}^{ab}\lambda_{\ell V}^{ab} D_a(q_2)D_b^*(q_1),
\nn
\end{eqnarray}     
and the axial-vector combinations can be obtained via the substitution
$A_j=V_j(V\rightarrow A)$.

The cross sections for the cases involving $\gamma \bar q$  
and the inverse configurations ($q \gamma $, $\bar q \gamma $) can be obtained in an analogous manner.
Next, we calculate the cross section of process (\ref{hh-mumu}) using 
the so-called {\it leading logarithm} (LL) approximation.

\section{Leading Logarithm Approximation}

\subsection{Photon-Quark Interaction}

Assume that the 4-momentum $p_5$  is proportional to $p_1$, such that
\begin{equation} 
p_5 = (1 - \eta) p_1,
\label{LL}
\end{equation}
i.e., the 3-momentum $\bf p_1$  is collinear with $\bf p_5$, 
indicating that the photon emission occurs parallel to the direction of the emitting quark.
Let us now introduce two useful combinations involving the parameter $\eta$:
\begin{equation}
\eta_\pm = 1 \pm \eta,
\label{eta_mp}
\end{equation}
so $p_5 = \eta_- p_1$.
Inserting equation~(\ref{LL}) into the momentum conservation law, we obtain:
\begin{equation} 
\eta p_1 + p_2 = p_3 + p_4.
\label{LL1}
\end{equation}     
With this result, the fundamental relations of collinear kinematics now take the following form:
\begin{equation} 
\eta s = s_1,\ \ \ \eta t = t_1,\ \ \ \eta u_1 = u.
\label{LL2}
\end{equation}

Our aim here is to derive the expressions for $S^V_q$ and $S_q^A$  
under the conditions specified by equation~(\ref{LL}).
After performing the necessary algebraic manipulations, we obtain:
\begin{eqnarray} 
S^V_{q,\rm LL} &=& 
  \frac{4}{w_1} P_{\gamma q} \eta (t^2 + u_1^2) 
+ \frac{4}{s}  \bigl( \eta \eta_- (t - u_1)^2 + 2 t u_1 \bigr),
\nn\\
S^A_{q,\rm LL} &=& 
  \frac{4}{w_1} P_{\gamma q} \eta (t^2 - u_1^2) 
+ \frac{4}{s}  \eta \eta_- (t^2 - u_1^2).
\label{S_LL}
\end{eqnarray}
The Altarelli--Parisi splitting function is evident in the initial terms of the expressions (\ref{S_LL}):
\begin{equation} 
P_{\gamma q} = \eta_-^2 + \eta^2.
\label{Pgq}
\end{equation}

In the framework of collinear kinematics (\ref{LL}), it is possible to elegantly express 
both all quantities $S_j^{V,A}$ and the phase space in terms of $\eta$. 
Employing equation (\ref{sqs}) and the ultrarelativistic approximation, 
we obtain the following relations in the partonic centre-of-mass system 
(where $\sqrt{s} = E_1+E_2$):
\begin{equation} 
E_{1,2} = \frac{M}{2\sqrt{\eta}},\ \ \ 
E_5 = \frac{M \eta_-}{2 \sqrt{\eta}}.
\label{LL_3}
\end{equation}  
Additionally, we require the differential expression
$ dE_5 = \frac{1}{4}M \eta_+ \eta^{-3/2} d\eta $.
To calculate $E_3$, $E_4$, and $\cal F$ -- 
quantities essential for determining the phase space -- we first extract the coefficients 
from equation (\ref{ABCDEF}):
$$ {\cal A} = E_5^2,\ \ \
   {\cal B} = E_5\cos\theta_3,\ \ \
   {\cal C} = \frac{M\eta_+}{2\sqrt{\eta}},
$$  
then
\begin{eqnarray} 
E_3 &=& \frac{M\sqrt{\eta}}{\eta_+ + \eta_- \cos\theta_3 },\ \ \
E_4 = \frac{M }{ 2\sqrt{\eta} } 
      \frac{ 1 + \eta^2 + \eta_-\eta_+ \cos\theta_3 }{ \eta_+ + \eta_- \cos\theta_3 },
\nn\\
{\cal F} &=& \frac{ ( \eta_+ + \eta_- \cos\theta_3)^2 }{ 1 + \eta^2 + \eta_-\eta_+ \cos\theta_3 }.
\label{E34F}
\end{eqnarray}  
The invariants (\ref{rad-invariants-qq}) exhibit the following properties: $w_1 \rightarrow 0,\ w_2 = \eta_- s$. 
Using these relations, the expressions can be derived from equation (\ref{x1-x2-R}):
\begin{eqnarray}
x_1 = \frac{M }{\sqrt{S}} \,  \frac{e^{+y}}{\eta},\ \ \
x_2 = \frac{M}{\sqrt{S}} \, e^{-y},\ \ \ 
J_x = \frac{4M}{S \eta_+ }.
\label{x1-x2-LL}
\end{eqnarray}
Finally, in the LL approximation, the invariants $t$ and $u$ take the following form:
\begin{equation} 
t = -\frac{M^2(1-\cos\theta_3)}{\eta_+ + \eta_- \cos\theta_3 },\ \ \
u = -\frac{M^2(1+\cos\theta_3)}{\eta_+ + \eta_- \cos\theta_3 }.
\label{tu-LL}
\end{equation}

Now that all invariants and the arguments of the parton distribution functions, $x_{1,2}$, 
are expressed solely in terms of $\eta$ and $\cos\theta_3$ (for fixed values of $M$ and $y$), 
we can proceed to integrate over the remaining phase-space variables: $\varphi_5$ and $\cos\theta_5$.
Owing to the simplicity of expression (\ref{S_LL}), it is sufficient to compute only a single integral:
\begin{eqnarray} 
  \int_{-1}^{1} d(\cos\theta_5) 
  \int_{0}^{2\pi} d\varphi_5 \frac{1}{w_1} &=& \frac{\pi}{E_1E_5} L_q,
\nn
\end{eqnarray}
where the {\it collinear logarithm} takes the following form:
\begin{equation} 
L_q 
= \log\frac{4E_5^2}{m_q^2}
= \log\frac{M^2 \eta_-^2}{m_q^2 \eta}.
\label{L-q}
\end{equation}  
For example, the integral
$$ \int_{-1}^{1} d(\cos\theta_5)   \int_{0}^{2\pi} d\varphi_5  = 4\pi $$
yields a constant contribution that is independent of $m_q$.

As a result of the LL approximation, the differential cross section assumes a simplified integral form:
\begin{eqnarray} 
d\sigma_{q}^{\rm LL} &=&
\frac{\alpha}{2\pi}
\sum\limits_{q} Q_q^2 
\int\limits_0^1 d\eta \,
f_\gamma^A(x_1) f_q^B(x_2) \, dx_1 dx_2 \times
\nn\\
&\times &
L_q P_{\gamma q} J_\eta d\sigma_{\bar q q}^{0}(\eta),
\label{xs-LL-q}
\end{eqnarray}     
where 
\begin{equation} 
J_\eta = \frac{2 \eta \eta_+ }{( \eta_+ + \eta_- \cos\theta_3)^2}.
\nn
\end{equation}  
This expression is in good agreement with the structure of formula (5) in \cite{ArSa}. 
It can be observed that, in formula (\ref{xs-LL-q}), the differential partonic 
Born cross section with ``shifted'' variables -- specifically, $t=t(\eta)$, $u_1 =u(\eta)/\eta$ 
(see formulas (\ref{tu-LL}) -- is factorized:
\begin{equation} 
d\sigma_{\bar q q}^{0}(\eta) =
\frac{\pi\alpha^2}{s} 
\bigl[ (t^2+u_1^2) V_{q} + (t^2-u_1^2) A_{q} \bigr]\, 
d(\cos\theta_3),
\end{equation}     
This corresponds to the cross section $d\sigma_{\bar q q}^{0} \equiv d\sigma_{\bar q q}^{0}(1)$ 
for $\bar q q$-annihilation into a dimuon:
$ \bar q(p_1) + q(p_2) \rightarrow \ell^-(p_3) + \ell^+(p_4)$.

\subsection{Photon-Lepton Interaction}

Let the 4-momentum $p_5$ be proportional to $p_2$:
\begin{equation} 
p_5 = \eta_- p_2,
\label{LL-l}
\end{equation}     
where $\bf p_2$ is collinear to $\bf p_5$. 
This configuration ensures the minimisation of $q_2^2$, 
which, in the case of photon exchange, maximises the photon propagator $D_{\gamma}(q_2)$.

Substituting (\ref{LL-l}) into the momentum conservation law yields:
\begin{equation} 
p_1 + \eta p_2 = p_3 + p_4,
\label{LL1-l}
\end{equation}     
whereupon the fundamental relations of collinear kinematics take the following form:
\begin{equation} 
s = s_1/\eta,\ \ \ t/\eta = t_1,\ \ \ u_1/\eta = u.
\label{LL2-2}
\end{equation}

To obtain $S^V_j$ and $S_j^A$ in the LL approximation, 
we need to perform a series of preliminary transformations.
We start by rewriting the momenta using the relations:
$p_2 \rightarrow q_2+p_5$, $p_1 \rightarrow p_3+p_4-q_2$.
Next, we apply approximations for scalar products that are valid in the LL regime:
$q_2p_5 \rightarrow -q_2^2/2$, $p_3p_4 \rightarrow q_2^2/2 + q_2p_2 $. 
It gives
\begin{eqnarray} 
S^V_{\ell} &=& 
 \Bigl( - 4 q_2^2 [ 2(q_2p_3)(p_3p_5) + (q_2p_3)^2 + 2(p_3p_5)^2 ] -
\nn\\ 
 &-& 8 m_q^2 (q_2p_3)^2 \Bigr)\frac{1}{(p_1p_3)(p_1p_4)} + (p_3 \leftrightarrow p_4),
\label{S_LL-l-1}
\\
S^A_{\ell} &=& 
 4 q_2^2 \frac{2(q_2p_3)(p_3p_5) + (q_2p_3)^2}{(p_1p_3)(p_1p_4)} - (p_3 \leftrightarrow p_4).
\nn
\end{eqnarray}
By analogy with the quark case, we obtain all the required expressions for the cross section; 
consequently, under the conditions specified in (\ref{LL-l}), we derive:
\begin{eqnarray} 
S^V_{\ell,\rm LL} &=& 
  \bigl( - 4 q_2^2 P_{\gamma q}/\eta - 8 m_q^2 \bigr) \Bigl( \frac{t}{u_1} + \frac{u_1}{t} \Bigr),
\nn\\
S^A_{\ell,\rm LL} &=& 
  4 q_2^2 (1 - 2/\eta) \Bigl( \frac{t}{u_1} - \frac{u_1}{t} \Bigr).
\label{S_LL-l}
\end{eqnarray}

The resulting expression is proportional to the corresponding Altarelli--Parisi splitting function:
\begin{equation} 
P_{\gamma \ell} = \frac{1 + \eta_-^2}{\eta}.
\end{equation}     
The arguments of the parton distribution functions are as follows:
\begin{eqnarray}
x_1 = \frac{M }{\sqrt{S}} \,  e^{+y},\ \ \
x_2 = \frac{M}{\sqrt{S}} \, \frac{e^{-y}}{\eta}.
\label{x1-x2-LL-l}
\end{eqnarray}

Once again, it is necessary to evaluate only a single integral:
\begin{eqnarray} 
  \int_{-1}^{1} d(\cos\theta_5) 
  \int_{0}^{2\pi} d\varphi_5 \frac{1}{q_2^2} &=& -\frac{\pi}{E_2E_5} L_\ell,
\nn
\end{eqnarray}
where the collinear logarithm now takes the following form:
\begin{equation} 
L_\ell 
= \log\frac{4E_2^2E_5^2}{m_q^2(E_2-E_5)^2}
= \log\frac{M^2 \eta_-^2}{m_q^2 \eta^3}.
\label{L-l}
\end{equation}  
As a result, the differential cross section in the LL approximation assumes an integral form:
\begin{eqnarray} 
d\sigma_{\ell}^{\rm LL} &=&
\frac{\alpha}{2\pi}
\sum\limits_{q} Q_q^2 
\int\limits_0^1 d\eta \,
f_\gamma^A(x_1) f_q^B(x_2) \, dx_1 dx_2 \times
\nn\\
&\times &
L_\ell P_{\gamma \ell} J'_\eta d\sigma_{\gamma\gamma}^{0}(\eta),
\label{xs-LL-l}
\end{eqnarray}     
where $J'_\eta = J_\eta/\eta^2$.
This expression is again in direct agreement with \cite{ArSa}. 
The differential partonic Born cross section, 
formulated in terms of the ``shifted'' variables in (\ref{xs-LL-l}), is factorized as follows:
\begin{equation} 
d\sigma_{\gamma\gamma}^{0}(\eta) =
\frac{\pi\alpha^2}{s} 
\frac{t^2+u_1^2}{t u_1} 
d(\cos\theta_3),
\end{equation}     
this corresponds to the cross section for $\gamma\gamma$-annihilation into a dimuon pair:
$\gamma(p_1) + \gamma(p_2) \rightarrow \ell^-(p_3) + \ell^+(p_4)$.

\section{Forward-Backward Asymmetry and the Relative Corrections}

Forward-backward asymmetry ($A_{\rm FB}$) 
is a crucial observable in high-energy physics experiments, 
including those at the LHC. 
It provides insights into the nature of particle interactions, the structure of the Standard Model, 
and potential signals of physics beyond the SM:
\begin{equation}
A_{\rm FB} = \frac{\sigma_{\rm F}-\sigma_{\rm B}}{\sigma_{\rm F}+\sigma_{\rm B}},
\label{afb}
\end{equation}
here, $\sigma_{\rm F}$ corresponds to the region $\cos\theta^* > 0$, 
and $\sigma_{\rm B}$ corresponds to $\cos\theta^* < 0$, where $\theta^*$ 
denotes the angle of dilepton emission in the Collins--Soper reference frame. 
The exact formula for $\theta^*$ is provided in \cite{Collins:1977iv}; 
in the notation adopted in the present paper, it takes the following form:
\begin{equation}
\cos\theta^*
=  
\frac{ {\rm sgn}[x_2(t+u_1)-x_1(t_1+u)] (tt_1 - uu_1) }{M\sqrt{s(u+t_1)(u_1+t)}}.
\label{cos-star-2}
\end{equation}

To obtain the total cross sections $\sigma_{\rm F,B}^{c}$ 
that contribute to the asymmetry, we must perform a numerical integration 
over the invariant lepton mass $M$ within the interval $M_k \leq M \leq M_l$. 
The standard mass intervals adopted for the high-energy region in the CMS experiment 
are defined by the following set of values (in TeV):
$M_{k,l} = \{0.2, 0.22, 0.243,$ $ 0.273, 0.32, 0.38,$ 
$0.44, 0.51, 0.6,$ $0.7,$ $0.83,$ $1.0,$ $1.2, 1.5, 2.0,$ $3.0, 6.5 \} $.

Integration over the remaining variables must be performed within 
a specific portion of the phase space $\Omega$, 
which corresponds to the designated detector acceptance region. 
This region is mathematically described by the factor $\Theta$ 
(for further details, see Sect.\ref{2-3} and \cite{55-Au}). 
This procedure yields the fiducial cross section:
\begin{equation}                                                                       
 \sigma_j^{c} = \int_\Omega d\sigma_j^{c}\cdot \Theta.
\label{tot-xs}
\end{equation}

\begin{figure}
    \centering
    \mbox{
        \includegraphics[width=0.39\textwidth]{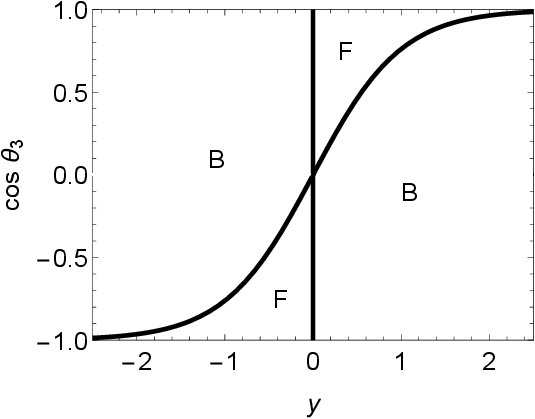}
         }
    \caption{
Regions of integration in variables $y$ and $\cos\theta_3$ 
for the forward ($\sigma_{\rm F}$) and backward ($\sigma_{\rm B}$) cross sections.
}
    \label{fig.FB}
\end{figure}

Usually, for the description of the process, we take into account the Born 
Drell--Yan contribution (denoted by the index $0$) 
and several additional contributions, which are marked by the index $c$ [see (\ref{c})].
Then, after some algebraic manipulations, the radiatively corrected forward-backward asymmetry 
takes the following form:
\begin{equation}                                                                       
A_{\rm FB}^c 
= 
\frac{\sigma_{\rm F}^0 + \sum_c\sigma_{\rm F}^c - \sigma_{\rm B}^0 - \sum_c\sigma_{\rm B}^c}
     {\sigma_{\rm F}^0 + \sum_c\sigma_{\rm F}^c + \sigma_{\rm B}^0 + \sum_c\sigma_{\rm F}^c}
= 
A_{\rm FB}^0
\frac{1 + \sum_c\delta_{-}^c}
     {1 + \sum_c\delta_{+}^c},
\label{afb-c}
\end{equation}
where two relative corrections, defined as
\begin{equation}                                                                       
\delta_\pm^{c} 
= \frac{ \sigma_{\rm F}^c \pm \sigma_{\rm B}^c }
       { \sigma_{\rm F}^0 \pm \sigma_{\rm B}^0 }
\label{rel-corr}
\end{equation}
possess a crucial property: they are additive. 
This means that to determine the net effect on the observable asymmetry 
from multiple sources of corrections (e.g., for ($c=a$ and $c=b$)), one should follow these steps:
\begin{enumerate}
\item
Calculate the relative corrections $\delta_{\pm}^a$ and $\delta_{\pm}^b$ 
separately for each source of corrections.
\item
Add the corrections linearly to obtain the combined relative correction:
$\delta_{\pm}^{a+b} = \delta_{\pm}^a + \delta_{\pm}^b$.
\item
Use the resulting combined correction $\delta_{\pm}^{a+b}$ 
in formula (\ref{afb-c}) to compute the net forward-backward asymmetry.
\end{enumerate}

\section{Quark Mass Singularity}

To address the issue of quark mass dependence (i.e., quark singularity, QS), we employ the  
$\overline{MS}$-subtraction technique. 
This involves subtracting a specific QS-term from the exact cross section given in (\ref{cs_hh-mumu-3}).
The QS-term is constructed to mimic the structure of the LL terms 
presented in (\ref{xs-LL-q}) and (\ref{xs-LL-l}), but with prescribed substitutions 
that account for the quark mass effects.
For the quark case
\begin{eqnarray} 
d\sigma_{q}^{\rm QS} &=&
\frac{\alpha}{2\pi}
\sum\limits_{q} Q_q^2 \log\frac{M^2}{m_q^2}
\int\limits_0^1 d\eta \, 
P_{\gamma q} \, J_\eta \, d\sigma_{\bar q q}^{0}(\eta) 
\times
\nn
\\
&\times &
f_\gamma^A(x_1) f_q^B(x_2) \, J_x \, dM dy
\label{xs-QS2-q}
\end{eqnarray}
and for the lepton case
\begin{eqnarray} 
d\sigma_{\ell}^{\rm QS} &=&
\frac{\alpha}{2\pi}
\sum\limits_{q} Q_q^2 
\int\limits_0^1 d\eta \,
P_{\gamma \ell} \Bigl(\log\frac{M^2}{m_q^2} - 2\log\eta -1 \Bigr)
\times
\nn
\\
&\times &
J'_\eta
d\sigma_{\gamma\gamma}^{0}(\eta)
f_\gamma^A(x_1) f_q^B(x_2) \, J_x \, dM dy.
\label{xs-QS2-l}
\end{eqnarray}     

As a result, the physical cross section takes the following form:
\begin{equation}                                                                       
d\sigma_{j}^{\rm phys} =
d\sigma_{j}^{\rm ex} - d\sigma_{j}^{\rm QS}, 
\label{xs-Phys}
\end{equation}
and now it does not depend on quark masses.

To ensure that the physical result does not depend on the unphysical 
value of the quark mass, let us examine the obtained numerical results presented in Table \ref{table1}.
The $M$ interval chosen, as indicated in the table caption 
(corresponding to large invariant dimuon mass), 
corresponds to the beginning of a critical region for NP searches.

\begin{table}
\centering
\caption{\label{table1} 
Relative corrections as a function of quark mass ($m_q=10^{n}M$) 
for integration over $M$ in the range from $1.0$ to $1.2$ TeV and 
over the full CMS rapidity interval ($|y| < 2.5$).
}
\begin{tabular}{cccc}
 \hline
 \hline
$n$   &$ \delta_+^{q, \rm ex}    $&$ \delta_+^{q, \rm QS} $&$  \delta_+^{q, \rm phys}$ \\
 \hline   
$-6$  &  0.001949 & 0.002095 &$ -0.000146 $   \\
$-5$  &  0.001602 & 0.001748 &$ -0.000146 $   \\
$-4$  &  0.001255 & 0.001400 &$ -0.000145 $   \\
$-3$  &  0.000908 & 0.001048 &$ -0.000140 $   \\
 \hline
$n$   &$ \delta_-^{q, \rm ex}    $&$ \delta_-^{q, \rm QS} $&$  \delta_-^{q, \rm phys}$ \\
 \hline   
$-6$  &  0.001511 & 0.001650 &$ -0.000138 $   \\
$-5$  &  0.001237 & 0.001376 &$ -0.000139 $   \\
$-4$  &  0.000965 & 0.001103 &$ -0.000138 $   \\
$-3$  &  0.000691 & 0.000826 &$ -0.000135 $   \\
 \hline
 \hline
$n$   &$ \delta_+^{\ell, \rm ex}    $&$ \delta_+^{\ell, \rm QS} $&$  \delta_+^{\ell, \rm phys}$ \\
 \hline
$-6$  &$ 0.111949  $&$ 0.114047   $&$ -0.002098 $   \\ 
$-5$  &$ 0.093314  $&$ 0.095375   $&$ -0.002061 $   \\ 
$-4$  &$ 0.074628  $&$ 0.076712   $&$ -0.002084 $   \\ 
$-3$  &$ 0.055960  $&$ 0.058047   $&$ -0.002087 $   \\ 
 \hline
$n$   &$ \delta_-^{\ell, \rm ex}  $&  $ \delta_-^{\ell, \rm QS} $&$  \delta_-^{\ell, \rm phys}$ \\
 \hline
$-6$  &$ 0.005078  $&  $ 0 $&$  $ 0.005078  \\
$-5$  &$ 0.005169  $&  $ 0 $&$  $ 0.005169  \\
$-4$  &$ 0.005145  $&  $ 0 $&$  $ 0.005145  \\
$-3$  &$ 0.005153  $&  $ 0 $&$  $ 0.005153  \\
\end{tabular}
\end{table}

We observe a good independence of the relative corrections with respect 
to the quark mass over a wide range spanning four orders of magnitude. 
The interference term exhibits no dependence on $m_q$  whatsoever.
An additional important remark is that $\delta_-^{\ell, \rm QS} =0$, 
as this quantity is purely of QED origin. 
For such quantities, the forward-backward difference ($\rm F-B$) yields exactly zero.

\section{Numerical Analisys }

The electroweak parameters and leptonic masses were obtained from \cite{PDG20}, 
while the parton distribution functions were taken from the latest NNPDF4.0 set \cite{NNPDF40}, 
with the natural scale choice $Q \equiv \sqrt{Q^2} = M$. 
For reliable results, it is necessary to use a new generation of PDFs, which are specifically 
designed to remain accurate and stable under conditions of arbitrarily large $Q^2$ values.
The LHAPDF 6.5.5 library \cite{LHAPDF6.5.5} was employed for the interface.

The prescriptions outlined in the CMS LHC Technical Design Report \cite{TDR} are as follows.
\begin{enumerate}
\item
The reaction referenced as (\ref{hh-mumu}) involves unpolarized protons with a total 
proton centre-of-mass energy of $\sqrt{S}=13.6\ \mbox{TeV}$ (Run 3/HL-LHC) 
and features a $\mu^-\mu^+$ final state.
\item
The CMS detector imposes standard kinematic restrictions on the angular 
acceptance of detected leptons. For a lepton $\ell^-$, 
these constraints are expressed either in terms of the polar angle as
$-\zeta^* \leq \cos\theta \leq \zeta^*$ or in terms of pseudorapidity as $|y(\ell)| \leq y(\ell)^*$,
where the pseudorapidity is defined as $y(\ell) = - \log\tan \frac{\theta}{2}$,
and the relation between $\cos\theta$ and $y(\ell)$ is given by $\cos\theta  = \tanh y(\ell)$.
For the CMS detector, the specific values of the acceptance parameters are:
$y(\ell)^* = 2.5$, $\zeta^* \approx 0.986614$.
Identical kinematic restrictions are applied to the positively charged lepton $\ell^+$.
\item
The standard restriction of the CMS detector on the transverse momentum components of particles:
${p_{3,4}}_T \geq 20\ \mbox{GeV}$.
\end{enumerate}

We require a set of kinematic expressions for energies, angles, and related variables 
in the hadron centre-of-mass system (denoted hereafter as h.c.m.s.). 
To derive these quantities, we first identify the corresponding expressions 
in the parton centre-of-mass system (p.c.m.s.) and subsequently apply 
the transformation rules specified in Eq.~(\ref{QPM}) to translate the results into the h.c.m.s.
In the p.c.m.s., the following relations hold:
\begin{eqnarray}
 t + u     &=& -2(p_1+p_2)p_3 = -4 E_1 E_3 = -2 \sqrt{s}\, E_3,
\nn
\\ 
 t_1 + u_1 &=& -2(p_1+p_2)p_4 = -4 E_1 E_4 = -2 \sqrt{s}\, E_4.
\nn
\end{eqnarray}
From these relations, we obtain the energies of the final-state particles:
\begin{eqnarray}
&& E_3 = - \frac{t + u}{2 \sqrt{s}},\ \ 
   E_4 = - \frac{t_1 + u_1}{2 \sqrt{s}}\ \ \mbox{(in p.c.m.s)},
\nn
\\
&& E_3 = - \frac{\hat t + \hat u}{2 \sqrt{\hat s}},\ \ 
   E_4 = - \frac{\hat t_1 + \hat u_1}{2 \sqrt{\hat s}}\ \ \mbox{(in h.c.m.s)}.
\nn
\label{energii}
\end{eqnarray}
Analogously, for the angles (all notations are defined in Fig.~\ref{fig.vecs}), 
we derive the following equations:
\begin{eqnarray}
&& t    = -2 E_1 E_3 (1-\cos\theta_3) = (t+u) (1-\cos\theta_3)/2,
\nn
\\ 
&& u_1  = -2 E_1 E_4 (1-\cos\theta_4) = (t_1+u_1)(1-\cos\theta_4)/2.
\nn
\end{eqnarray}
Solving them, we obtain the following expressions for the cosines of the scattering angles:
\begin{eqnarray}
&& \cos\theta_3 = \frac{u-t}{u+t},\ \ 
   \cos\theta_4 = \frac{t_1-u_1}{t_1+u_1}\ \ \mbox{(in p.c.m.s.)},
\nn
\\
&&\cos\theta_3 = \frac{\hat u - \hat t}{\hat u + \hat t},\ \ 
  \cos\theta_4 = \frac{\hat t_1 - \hat u_1}{\hat t_1 + \hat u_1}\ \ \mbox{(in h.c.m.s.)}.
\nn
\end{eqnarray}

For the transverse and longitudinal components of the 3-momenta in the hadronic center-of-mass system, 
the following expressions hold:
\begin{eqnarray}
&& {p_3}_T = E_3 \sin\theta_3 = \sqrt{\hat t\hat u /\hat s},\ \ 
   {p_4}_T = E_4 \sin\theta_4 = \sqrt{\hat t_1\hat u_1/\hat s},
\nn
\\
&& {p_3}_z = E_3 \cos\theta_3 = \frac{\hat t - \hat u}{2 \sqrt{\hat s}},\ \ 
   {p_4}_z = E_4 \cos\theta_4 = \frac{\hat u_1 - \hat t_1}{2 \sqrt{\hat s}}.
\nn
\end{eqnarray}
For the pair rapidity $y$ in the h.c.m.s., we consider the total energy $E = E_3 + E_4$  
and the $z$-component of the pair momentum:
\begin{equation}
p_z = E_3 \cos\theta_3 + E_4 \cos\theta_4 
= \frac{1}{2\sqrt{\hat s}} (\hat t - \hat u - \hat t_1 + \hat u_1).
\nn
\end{equation}
Based on the expressions derived above, we obtain the following expression for the pair rapidity $y$:
\begin{eqnarray}
y = \frac{1}{2} \log\frac{E+p_z}{E-p_z}
= \frac{1}{2} \log\frac{\hat t_1 + \hat u}{\hat t + \hat u_1}.
\nn
\end{eqnarray}
Knowing $p_z$, we can determine the transverse momentum $p_T$ as follows:
\begin{eqnarray}
p_T^2 &=& |{\bf p_3 + p_4}|^2 - p_z^2  
= {(\hat t + \hat u_1)(\hat t_1 + \hat u)}/{\hat s} - M^2.
\nn
\end{eqnarray}

Finally, for the quantities required in the calculation of the forward-backward asymmetry, 
we obtain the following expressions:
\begin{eqnarray}
\cos\theta^*
= \frac{ 2 \, {\rm sgn}[p_z] }{M\sqrt{M^2+p_T^2}}
\bigl[ p^+(l^-) p^-(l^+) - p^-(l^-) p^+(l^+) \bigr],
\nn
\end{eqnarray}
where
\begin{eqnarray}
p^\pm(l^-) = \frac{1}{\sqrt{2}}(E_3 \pm {p_3}_z),\ \ \
p^\pm(l^+) = \frac{1}{\sqrt{2}}(E_4 \pm {p_4}_z).
\nn
\end{eqnarray}
After some simplifications in the h.c.m.s., the expressions for $\cos\theta^*$ take the following form:
\begin{equation}
\cos\theta^*
=  
2 \, {\rm sgn}[y] \frac{ E_3E_4 }{M E_\theta^*}
(\cos\theta_3-\cos\theta_4),
\nn
\end{equation}
where 
$$
E_\theta^* = {\sqrt{ E_3^2\sin^2\theta_3 + E_4^2\sin^2\theta_4 +2E_3E_4(1-\cos\theta_3\cos\theta_4)  }}.$$
We have used here the kinematic property relating rapidity and momentum components: $\tanh y = p_z/E$.

The boundary between 
the forward region (F-region, defined by $\cos\theta^* > 0 $, 
where ``F'' stands from ``forward'') and 
the backward region (B-region, defined by $\cos\theta^* < 0 $,
where ``B'' stands for ``backward'') -- as illustrated in Fig.~\ref{fig.FB} -- 
is characterised by the following conditions:
$y=0$ and $ \cos\theta_3 - \cos\theta_4 =0 $
The second condition is equivalent to the relation $\cos\theta_3 = \tanh y$.

\begin{figure}
    \centering
    \mbox{
        {\includegraphics[width=0.39\textwidth]{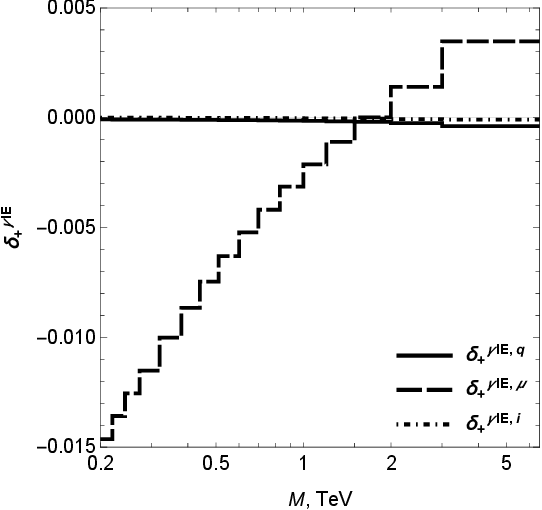}\label{p6}}
         }
    \caption{
Relative corrections $\delta_+$ induced by photon IE 
for different contributions: quarks, leptons (muons), and interference terms.
}
    \label{fig.rc6p}
\end{figure}

\begin{figure}
    \centering
    \mbox{
        {\includegraphics[width=0.39\textwidth]{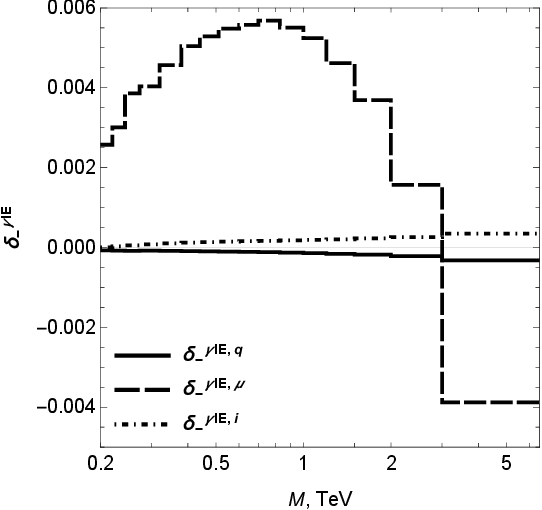}\label{m6}}        
         }
    \caption{
Relative corrections $\delta_-$ induced by photon IE 
for different contributions: quarks, leptons (muons), and interference terms.
}
    \label{fig.rc6m}
\end{figure}

Figures~\ref{fig.rc6p} and~\ref{fig.rc6m} illustrate the behaviour of the relative corrections, 
partitioned into their constituent contributions (quark channel, lepton channel, and their interference term), 
as a function of the invariant mass $M$.
Only the relative corrections for the $\mu$-case are significant in magnitude; $\delta_+^\mu$ 
is negative across the entire domain, whereas $\delta_-^\mu$ exhibits notable values 
only at large $M$; as $M$ increases, the absolute values of both quantities grow rapidly.
The relative corrections for the $q$-case and the interference case are small across 
the entire domain, with the sole exception of a gradual increase in magnitude as $M$ increases.
A similar behaviour is observed for the gluon IE \cite{60-Au},
indeed, the gluon case produces the same qualitative effects, 
the corresponding numerical values are markedly higher, which can be attributed to two reasons.
First, the strength of the gluon-quark interaction exceeds that of the photon-quark interaction. 
This is evident from the well-known formula describing the transition from the QED case 
to the QCD case for the cross section:
\begin{equation}
Q_q^2  \alpha 
\rightarrow  
\sum\limits_{a=1}^{N^2-1}  t^at^a \alpha_s  = \frac{N^2-1}{2N} I \alpha_s
\rightarrow 
\frac{4}{3} \alpha_s,
\label{zamena-QCD}
\end{equation}
where $\alpha$ is the fine-structure constant;
$\alpha_s$  is the strong coupling constant;
$2 t^a$ are the Gell-Mann matrices;
and $N=3$ is the quark generation number.
Second, the probability of finding a gluon within the proton 
is much greater than that of finding a photon. 
Consequently, the contribution of gluon-initiated processes 
to the overall cross section is significantly enhanced compared 
to any hypothetical photon-initiated ones: 
$$f_g^p(x,Q^2) \gg f_\gamma^p(x,Q^2).$$

\begin{figure}
    \centering
    \mbox{
        {\includegraphics[width=0.39\textwidth]{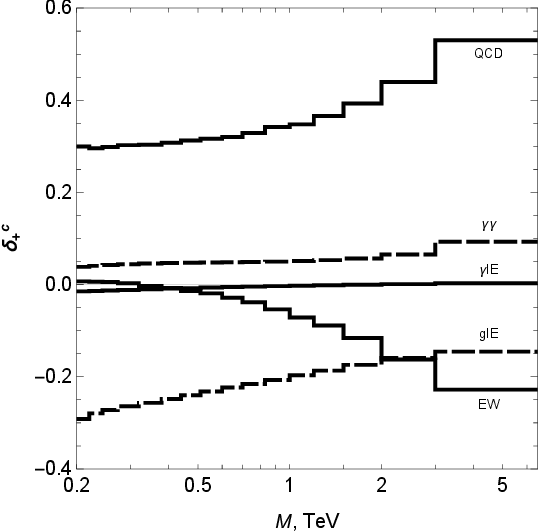}\label{p-all}}
         }
    \caption{
Relative corrections $\delta_+$ induced by various dimuon production mechanisms: 
Drell--Yan (electroweak and QCD), two-photon fusion, gluon and photon IE.
}
    \label{del-p}
\end{figure}
\begin{figure}
    \centering
    \mbox{
        {\includegraphics[width=0.39\textwidth]{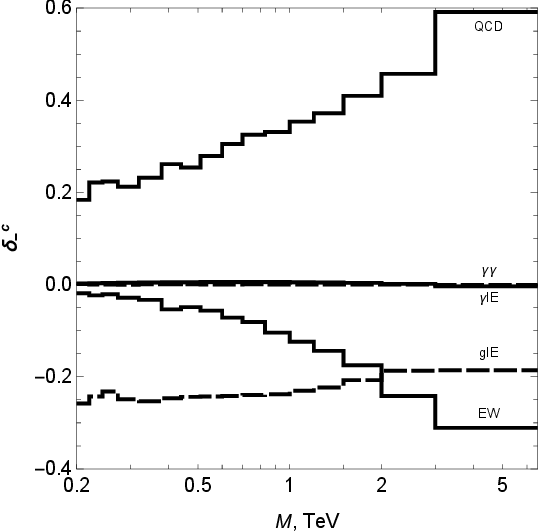}\label{m-all}}        
         }
    \caption{
Relative corrections $\delta_-$ induced by various dimuon production mechanisms: 
Drell--Yan (electroweak and QCD), two-photon fusion, gluon and photon IE.
}
    \label{del-m}
\end{figure}

Figures~\ref{del-p} and~\ref{del-m} present the relative corrections 
for the following processes: the electroweak RCs to the Drell--Yan process, 
the Drell--Yan process with QCD RCs, 
electroweak RCs with two-photon fusion, 
gluon IE, and photon IE plotted as a function of $M$ under CMS experiment conditions at the LHC, 
for the standard interval of dilepton rapidities  $|y| < 2.5$.
We observe that the obtained results for photon IE yield noticeable contributions only 
in the region of extremely large $M$, specifically for $M > 3$ TeV.

\begin{figure}
    \centering
    \mbox{
        \includegraphics[width=0.39\textwidth]{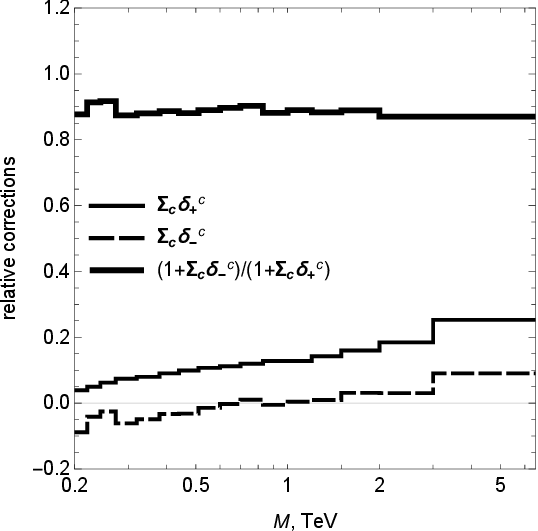}\label{del-all}
         }
    \caption{
Total relative corrections to the forward-backward asymmetry  $\delta_\pm$, 
and their multiplicative combination for Born asymmetry correction.
}
    \label{fig.rcA}
\end{figure}

Fig.~\ref{fig.rcA} shows the behaviour of the total relative corrections $\delta_\pm$ 
to the forward-backward asymmetry in dimuon production as a function of $M$ 
under CMS experiment conditions at the LHC. 
The figure also displays the multiplicative factor required to correct the Born asymmetry. 
Notably, multiple contributions exhibit significant mutual compensation -- 
a characteristic feature of such a complex observable as the forward-backward asymmetry. 
The use of additive $\delta_\pm$  corrections effectively resolves this problem.

\begin{figure}
    \centering
    \mbox{
        \includegraphics[width=0.39\textwidth]{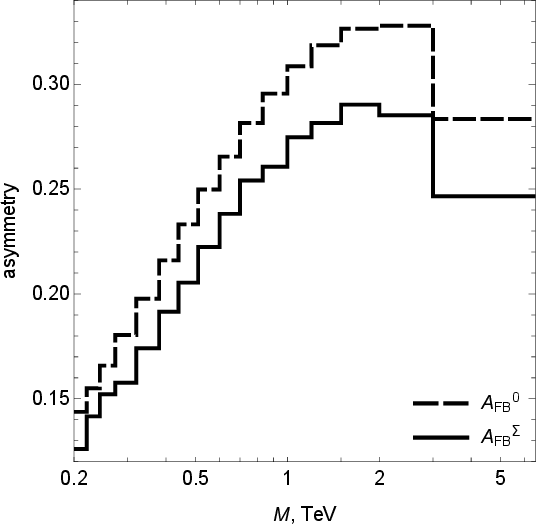}\label{A-all}
         }
    \caption{
Forward-backward asymmetry: comparison of the Born-level prediction and 
results including RCs, plotted as a function of $M$.
}
    \label{fig.A}
\end{figure}

Finally, Fig.~\ref{fig.A} illustrates the forward-backward asymmetry in dimuon production 
as a function of $M$ under CMS experiment conditions at the LHC.
The upper line represents the Born asymmetry, while the lower line 
corresponds to the asymmetry calculated with allowance for all currently 
studied mechanisms: radiative corrections in the Drell--Yan process, 
photon-photon fusion, and boson (gluon and photon) inverse emission.
The asymmetries accounting for radiative corrections differ significantly from 
the Born asymmetry at large values of $M$ espesially.

\section{Conclusion}

A detailed analysis of the forward-backward asymmetry in dilepton production 
through the photon inverse emission channel in hadron collisions is presented. 
The theoretically obtained results are of particular importance for the upcoming 
CMS LHC experimental program, which will focus on ultra-high energies and 
dilepton invariant masses exceeding 3 TeV (this kinematic region corresponds to the Run 3/HL-LHC regime).
The predicted effects are found to be at the 1 \% level, 
which corresponds to the expected statistical and systematic uncertainties.

\section{Aknowlegments}

The author is grateful to colleagues from the RDMS group of the CMS CERN Collaboration, 
and to A.\,B.~Arbuzov, Yu.\,M.~Bystritskiy, and S.\,V.~Shmatov for valuable discussions. 
The author also thanks CERN (CMS Group) for warm hospitality during research visits. 
This work was partly supported by the Convergence-2025 Research Program of the Republic of Belarus 
(Subprogram ``Microscopic World and Universe'').


\medskip
\begin{thebibliography}{99}

  \bibitem {CERN-W}
UA1 Collab. (G.~Arnison {\it et al.}), Phys. Lett. B  {\bf 122}, 103 (1983);
UA2 Collab. (M.~Banner {\it et al.}), Phys. Lett. B {\bf 122}, 476 (1983).
  \bibitem {CERN-Z}
UA1 Collab. (G. Arnison {\it et al.}), Phys. Lett. B {\bf 126}, 398 (1983);
UA2 Collab. (P.~Bagnaia {\it et al.}), Phys. Lett. B {\bf 129}, 130 (1983).
  \bibitem {SUSY}
J. Wess and B. Zumino, Phys. Lett. B {\bf 49}, 52 (1974).
  \bibitem {M-theory}
E. Witten, Nucl. Phys. B {\bf 463}, 383 (1996) [hep-th/9512219].
  \bibitem {Dark-matter}
G. Bertone, D. Hooper, and J. Silk, Phys. Rep. {\bf 405}, 279 (2005) [hep-ph/0404175].
  \bibitem {Axions}
M. Dine, W. Fischler, and M. Srednicki, Phys. Lett. B {\bf 104}, 199 (1981). 
  \bibitem {FIP}
P. Agrawal {\it et al.},
Eur. Phys. J. C {\bf 81}, 1015 (2021) [arXiv: 2102.12143 [hep-ph]].
  \bibitem {DY2}    
S.\,D.~Drell and T.\,-M.~Yan, Phys. Rev. Lett. {\bf 25}, 316, 902 (Erratum) (1970). 
  \bibitem {DY2-2}  
S.\,D.~Drell and T.\,-M.~Yan, Ann. Phys. (N.Y.) {\bf 66}, 578 (1971).  
  \bibitem {50-Au}  
V.\,A. Zykunov, Phys. At. Nucl. {\bf 84}, 492 (2021).
  \bibitem {55-Au} 
V.\,A. Zykunov, Phys. At. Nucl. {\bf 85}, 500  (2022).
  \bibitem {56-Au} 
V.\,A. Zykunov, Phys. At. Nucl. {\bf 86}, 9  (2023).
  \bibitem {22-Au}  
V.\,A. Zykunov, Phys. At. Nucl. {\bf 74}, 72 (2011).
  \bibitem {60-Au} 
V.\,A. Zykunov, Phys. At. Nucl. {\bf 88}, 81  (2025).
  \bibitem {BSH86} 
M. B\"ohm, H. Spiesberger, and W. Hollik, Fortschr. Phys. {\bf 34}, 687 (1986).
  \bibitem{Kuipers:2012rf} 
J.~Kuipers, T.~Ueda, J.~A.~M.~Vermaseren, and J.~Vollinga, 
Comput. Phys. Commun.  {\bf 184}, 1453 (2013) [arXiv: 1203.6543 [cs.SC]].
  \bibitem{FeynCalc} 
V. Shtabovenko, R. Mertig and F. Orellana, Comput. Phys. Commun. {\bf 256} (2020) 107478, arXiv:2001.04407.
  \bibitem{ArSa} 
A.\,B.~Arbuzov and R.\,R.~Sadykov, 
JETP {\bf 106}, 488 (2008) [arXiv: 0707.0423 [hep-ph]].
  \bibitem {Collins:1977iv}  
John C. Collins and Davison E. Soper, Phys. Rev. D {\bf 16}, 2219 (1977).
  \bibitem {PDG20} 
Particle Data Group (P.\,A.~Zyla {\it et al.}), Prog. Theor. Exp. Phys. {\bf 2020}, 083C01 (2020).
  \bibitem {NNPDF40} 
NNPDF Collaboration (R.\,D.~Ball {\it et al.}), 
Eur. Phys. J. C {\bf 84}, 540 (2024) [arXiv:2401.08749 [hep-ph]].
  \bibitem {LHAPDF6.5.5} 
A. Buckley {\it et al.},
Eur. Phys. J. C {\bf 75}, 132  (2015) [arxiv:1412.7420 [hep-ph]].
  \bibitem {TDR} 
CMS Collab. (G.\,L. Bayatian {\it et al.}), J.~Phys. G {\bf 34}, 995 (2007).
%
\end{thebibliography}
\end{document}